\documentclass[letterpaper,nofootinbib,showpac,onecolumn,eqsecnum,superscriptaddress,floatfix]{revtex4}

\usepackage{amsthm}
\usepackage{amssymb}   
\usepackage{dsfont}
\usepackage{mathtools} 
\usepackage{hyperref}
\hypersetup{colorlinks=true,linkcolor=blue,urlcolor=blue,citecolor=blue}
\usepackage{accents}
\usepackage{tensor}
\usepackage{xcolor}
\usepackage{graphicx}
\usepackage[cal=boondox]{mathalfa}
\usepackage{lipsum}
\usepackage{relsize}
\usepackage{color}
\usepackage{comment}
\usepackage{nameref}
\usepackage{lineno} 
\usepackage{orcidlink}
\usepackage{ulem}
\usepackage{soul}

\usepackage{natbib}
\begin{document}
%%%%%%%%%%%%%%%%%%%%%%%%%%%%%%%%%%%%%%%%%%%%%

% \title{Background Field theory in a quintom cosmological model---Phantom Cosmology from BF-theory and Topological Fluid}

\title{$BF$ effective action topological implications}

 \author{Erika Amador-Ramos \orcidlink{0009-0000-6079-1682}}
 \email[]{erika.amador8815@alumnos.udg.mx}
\email[]{aldo.martinez@academicos.udg.mx}  
\author{Aldo Mart\'inez-Merino\orcidlink{0000-0003-4767-6879}}
\email[Corresponding author:]{roberto.santos@academicos.udg.mx}
\author{Roberto Santos-Silva\orcidlink{0000-0002-1265-0938}}
\email[]{aldo.martinez@academicos.udg.mx}
\affiliation{Departamento de Ciencias Naturales y Exactas, CUValles,
Universidad de Guadalajara,
C.P. 46600, Ameca, Jalisco, M\'exico}

\date{\today}
	
\begin{abstract}
%

% %  
%
We investigate the effective action of a 7-dimensional topological $BF$ theory under a 4-dimensional reduction. We discuss the topological 
implications of the resulting theory and analyze the corresponding field equations. In particular, we derive exact solutions to these equations of 
motion in both Minkowski and Schwarzschild spacetimes.

\end{abstract}
	
\maketitle
	
%%%%%%%%%%%%%%%%%%%%%%%%%%%%%%%%%%%%%%%%%%%%%%%%%%%%
\section{Introduction}
%%%%%%%%%%%%%%%%%%%%%%%%%%%%%%%%%%%%%%%%%%%%%%%%%%%%%

In this article we investigate some solutions of the $7$-dimensional model proposed in \cite{MEHM}, which is formulated as a topological $BF$ theory with a metric-independent Lagrangian. It is well known that an $n$-dimensional $BF$ theory is a diffeomorphism-invariant gauge theory with an action given by
\begin{equation}
    S[A,B]=\int_{M}\mathrm{Tr}(B\wedge F[A]),
\end{equation}
thereby defining a topological field theory \citep{Horowitz, BBRT, CGRS, CCRR}.

$BF$ theory was originally introduced in the $4$-dimensional case by Plebański in 1977 \citep{Plebanski}. With an appropriate modification, 
four-dimensional General Relativity (GR) can be recast as a $BF$ theory. This is accomplished by imposing an additional constraint on 
the $B$ field, commonly referred to as the simplicity constraint. Solving this constraint yields a self-dual formulation of GR and endows the 
theory with local degrees of freedom (see also \cite{CMPR, CGM, PF}). From this perspective, the $B$ field may be regarded as the 
fundamental variable, rather than the spacetime metric $g_{\mu\nu}$ \cite{CJD}. Furthermore, in dimensions $n \geq 4$ and in the absence 
of a cosmological constant $\Lambda$, GR can likewise be formulated as a constrained BF theory \cite{FKP, MEC}.

The original proposal by \cite{MEHM} frames this 7-dimensional $BF$ theory within type IIB string theory, as it includes the potential for 
an extended object in addition to the gauge field $A$. Another theory defined on a 7-dimensional 
manifold worth mentioning is Hitchin's functional \cite{Hitchin_1, Hitchin_2}-in particular, on a $G_2$ manifold-which 
under dimensional reduction, yields a series of gravity theories depending on the dimensionality of the reduced manifold. 
The main difference is that the latter model relies on the existence of a 3-form that defines the manifold.

The implications of the 7-dimensional $BF$ theory studied in the present paper for cosmological models have been investigated in 
\cite{ESMH_1}, while its canonical analysis was carried out in \cite{ESMH_2}.

This work begins with a brief review of the original higher-dimensional theory. We then perform a dimensional reduction to recover the 
$4$-dimensional model that forms the focus of our analysis. For completeness, Appendix A summarizes the essential definitions 
and conventions on differential forms used throughout the paper. The explicit functional forms of the solutions found for the Schwarzschild 
background are deferred to the remaining appendices. Unless otherwise stated, we adopt natural units where $\hbar = c = 1$.

%%%%%%%%%%%%%%%%%%%%%%%%%%%%%%%%%%%%

%%%%%%%%%%%%%%%%%%%%%%%%%%%%%%%%%%%%%%%%%%%%%%%%%%%%%%%%%%%%%%%%%%%%%%%%%%%%%%%%%%%%%%%%%%%%%%%%%%%%%%%%%%%%%%%%%%%%%%%%%%%%%%%%%%%%%%%%%%%
\section{$4$-dimensional $BF$ effective action}\label{4dim}
%%%%%%%%%%%%%%%%%%%%%%%%%%%%%%%%%%%%%%%%%%%%%%%%%%%%%%%%%%%%%%%%%%%%%%%%%%%%%%%%%%%%%%%%%%%%%%%%%%%%%%%%%%%%%%%%%%%%%%%%%%%%%%%%%%%%%%%%%%%

In \cite{MEHM, EHMM}, the authors introduced a $7$-dimensional BF theory defined by the action
\begin{equation}
    S_7 = m_0 k \int_{M_7} B_7 \wedge F_7,
    \label{action}
\end{equation}
where $m_0$ denotes a mass scale parameter and $k$ is the coupling constant of the theory. The $7$-dimensional manifold $M_7$ is taken to 
be a product manifold of the form $M_7=M_3 \times M_4$, where $M_3$ is a nontrivial compact manifold consisting of Seifert spheres glued to a 
$2$-torus, and $M_4$ represents a four-dimensional spacetime.

The fields $B_7 \in \Omega^3 (M_7, \mathbb{R}) \otimes \mathfrak{u}(1)$ and $F_7 \in \Omega^4 (M_7, \mathbb{R}) \otimes \mathfrak{u}(1)$ 
are abelian gauge fields; $\Omega^p (M_7, \mathbb{R})$ denotes the set of real differential $p$-forms in $M_7$. In particular, $F_7$ is the field 
strength associated with a $3$-form gauge potential $A_7$, i.e., $F_7=dA_7$.

In order to perform a dimensional reduction, we consider differential forms $\alpha \in H^2 (M_3, \mathbb{R})$ and $\beta \in H^1(M_3, \mathbb{R})$ 
belonging to the cohomology groups of $M_3$, and adopt the ansatz
\begin{equation}
F_7 = \alpha \wedge F, \qquad B_7 = \beta \wedge B,
\end{equation}
with $B, F \in \Omega^2 (M_4, \mathbb{R}) \otimes \mathfrak{u} (1)$ abelian gauge fields defined on $M_4$, with $F=dA$ for a $1$-form 
gauge potential $A$; in such dimensional reduction, the $B$-field is now at the same time a gauge potential for a field strength $H$, this is
$H = dB$. The resulting effective $4$-dimensional theory is described by an action whose explicit form is given by 
\cite{EHMBL}\footnote{Originally, the constant $k$ is a matrix related to the linking number between $1$-homology groups in the $3$-dimensional manifold (before compactification).} 
\begin{equation} \label{faction}
    S[A, B, g] = \int_{M_4} \left[ m_0 k A \wedge dB - \frac{1}{2} dA\wedge \star dA - \frac{1}{2} dB \wedge \star dB - \frac{1}{4m_0 k} \star dA 
    \wedge d \star dB \right],
\end{equation}
where $g$ is the metric in the manifold $M_4$. This action is invariant under large gauge transformations, transformations 
of the form
\begin{subequations}
\begin{eqnarray} \label{gaugeA}
A  \quad & \rightarrow & \quad A - \tilde{a} \\
%\frac{1}{m_0 k}d a,
\label{gaugeB} B \quad & \rightarrow & \quad B + \tilde{b}
%\frac{1}{m_0 k}d  b. 
\end{eqnarray}
\end{subequations}
where $\tilde{a} \in H^1 (M_4, \mathbb{Z})$ and $\tilde{b} \in H^2 (M_4, \mathbb{Z})$ are 1- and 2-forms, respectively, 
pertaining to the corresponding cohomology groups of the 4-dimensional manifold. Observe that, after dimensional 
reduction, the theory defined by the action (\ref{faction}) is no longer topological, since it depends explicitly on the 
metric of the manifold $M_4$ through the appearance of the Hodge star operator \cite{EHMBL}.

We next rewrite the effective action in eq. \eqref{faction} using the inner product of $p$-forms, which is defined in terms of the wedge product 
and the Hodge star operator, $\star$. This formulation proves more convenient for the subsequent analysis, in particular for deriving and 
analyzing the equations of motion (eom). Explicit definitions and conventions are collected in Appendix \ref{Appendix_A}. First, we have the
global inner product
\begin{equation}
(\eta, \omega) := \int_{M_4} \eta \wedge \star \omega.
\end{equation}\label{effective_action}
Then the effective action in Eq. \eqref{faction} can be recast in the form
\begin{equation}
S[A, B, g] = m_0 k (A, \star dB) - \frac{1}{2} (dA, dA) - \frac{1}{2} (dB,dB) - \frac{1}{4m_0 k} (d\star dB, dA).
\end{equation}
In order to perform the variation wrt the $A$ field, it is convenient to rewrite the last term of the preceding equation using standard identities for 
differential forms. In this way, the action takes the equivalent form
\begin{equation}\label{action_rewritten}
    S[A, B, g] = m_0 k (A, \star dB) - \frac{1}{2} (A, d^{\dagger} dA) - \frac{1}{2} (B, d^{\dagger}dB) - \frac{1}{4m_0 k} (A, d^{\dagger} d \star dB),
\end{equation}
where $d^\dagger$ denotes the adjoint of the exterior derivative $d$, (for conventions see Appendix \ref{Appendix_A}.
%\eqref{dAdjoint}
Hence, performing such variation we obtain
\begin{equation}
    \delta_A S = \left( \delta_A A, m_0 k \star dB - d^{\dagger} dA - \frac{1}{4m_0 k} d^{\dagger} d \star dB \right) = 0,
\end{equation}
yielding the following equation of motion (eom)
\begin{equation}\label{EOM_1}
(d^{\dagger}d - m^2) \star dB = -2 md^\dagger dA.
\end{equation}
In terms of the corresponding field strengths, which is also a closed form, $d H = 0$, and by the definition of the adjoint 
$d^\dagger$, we arrive at the first eom
\begin{equation}\label{EOM_H}
    (\square - m^2) H = -2m\; d \star F,
\end{equation}
where $\Box = d^\dagger d + d d^\dagger$  denotes the laplacian\footnote{Is possible to add the term $dd^{\dagger}$, since this operator 
acting on $\star dB$ vanishes.}; $m = 2 m_0 k$ is the topological mass.

A similar procedure is followed to calculate the variation of the action wrt the $B$ field. For this purpose, it is convenient to express the first and 
last terms appearing in eq. \eqref{action_rewritten} using the following property,
\begin{equation}
m_0 k \int_{M_4}A \wedge dB = -m_0 k (B, \star dA) \qquad \Rightarrow \qquad
(d\star dB, dA) = -(B, d^{\dagger}d\star dA) \,.
\end{equation}
Then, the effective action is rewritten as
\begin{equation}
S[A, B, g] = -m_0 k (B, \star dA)-\frac{1}{2}(dA, dA)-\frac{1}{2}(B, d^{\dagger}dB)+\frac{1}{4m_0 k}(B, d^{\dagger}d\star dA).
\end{equation}
Now, the variation with respect to the field $B$ reads
\begin{equation}
\delta_B S = \left( \delta B, -m_0 k \star dA - d^{\dagger} dB + \frac{1}{4m_0 k} d^{\dagger} d \star dA \right) = 0,
\end{equation}
yielding the eom
\begin{equation}\label{EOM_2}
(d^{\dagger}d - m^2)\star dA = 2 m\; d^\dagger dB.
\end{equation}
Again, we have that this equation is recast in the form
\begin{equation}\label{EOM_F}
    (\square - m^2) F = 2m\; d \star H,
\end{equation}
as is reported in \cite{EHMBL}; it is assumed at the same time that $d F = 0$.

We have then a system of Proca-like fields with the same (topological) mass whose eom's \eqref{EOM_H} and \eqref{EOM_F} are coupled with one
another. We conciliate the fact that these gauge fields have now mass arguing that the gauge transformations are not connected
to the identity element, as we said earlier these are large gauge transformations, plus, we are generating this mass by
topological means. In the next section we further analyze this system in order to find solutions within a given metric.

We finalize this section noting that the equations \eqref{EOM_1}  and \eqref{EOM_2} are invariant under the gauge transformations \eqref{gaugeA} 
and \eqref{gaugeB}, as a direct calculation can attest.

% For Original Research articles, please note that the Material and Methods section can be placed in any of the following ways: before Results, before Discussion or after Discussion.

%%%%%%%%%%%%%%%%%%%%%%%%%%%%%%%%%%%%%%%%%%%%%%%%%%%%%%%%%%%%%%%%%%%%%%%%%%%%%%%%%%%%%%%%%%%%%%%%%%%%%%%%%%%%%%%%%%%%%%%%%%%%%%%%%%%%%%%%%%%
\section{Equations of motion and the spacetime}
\label{EOM_and_ST}
%%%%%%%%%%%%%%%%%%%%%%%%%%%%%%%%%%%%%%%%%%%%%%%%%%%%%%%%%%%%%%%%%%%%%%%%%%%%%%%%%%%%%%%%%%%%%%%%%%%%%%%%%%%%%%%%%%%%%%%%%%%%%%%%%%%%%%%%%%%

In this section, we further analyze the equations of motion and their relation with the topological structure of the spacetime
manifold. We begin by considering eq.~\eqref{EOM_1}
\begin{eqnarray}
(d^\dagger d - m^2) \star d B = - 2m d^\dagger d A,
\end{eqnarray}
which we can rewritten by factoring out $\star d$, leading to
\begin{equation}\label{eq_1}
    \star d \left( [d^\dagger d - m^2] B + 2m \star d A \right) = 0,
\end{equation}
This implies that the form in parenthesis is a closed 2-form, say $\alpha \in H^2 (M_4, \mathbb{Z})$. Invoking the gauge transformation (\ref{gaugeA}), we can absorb $\alpha$ by suitable gauge transformation. Hence, we arrive at
\begin{equation}\label{eom_A2}
(d^\dagger d - m^2) B = -2m \star dA,
%\qquad \mbox{or} \qquad ( \square - m^2) B = -2m \star F
\end{equation}

In a similar fashion, we can rewrite eq. \eqref{EOM_2} to take the form
\begin{equation}
\star d \left( [d^\dagger d - m^2] A - 2m \star d B \right)=0.
\end{equation}
This implies again that this is an element of the first cohomology group $H^1(M_4, \mathbb{Z})$. Thus it is equal to a closed 
1-form, say $\beta$, which by a gauge transformation \eqref{gaugeB} can be absorbed. Finally, arriving at the equation
\begin{equation}\label{eom_B2}
(d^\dagger d - m^2) A = 2m \star dB
%\qquad \mbox{or} \qquad (\square - m^2) A = 2m \star H.
\end{equation}

In this way we observe that in general solving the eom's we can compute explicitly\footnote{Either trivial or not trivial 
cohomology elements.} the cohomology groups $H^1(M_4,\mathbb{Z})$ and $H^2(M_4,\mathbb{Z})$, and by duality we 
find the $H^3(M_4,\mathbb{Z})$ (through Hodge duality)\footnote{$dim H^{p}(M,\mathbb{Z})=dimH^{n-p}(M,\mathbb{Z})$, 
where $p$ is the $p$-th cohomology group and $n$ is the dimension of the manifold $M$.}.

In order to find solutions to these equations, we look for a way to decouple these same equations. Applying the operator
$d^\dagger d - m^2$ for instance, to \eqref{eom_B2} taking into account \eqref{eom_A2}, explicitly we have for the lhs
\begin{eqnarray}
(d^\dagger d - m^2) (d^\dagger d - m^2) A &=& ( d^\dagger d \cdot d^\dagger d - 2 m^2 d^\dagger d + (m^2)^2) A,
\end{eqnarray}
meanwhile for the rhs we have
\begin{eqnarray}
2m (d^\dagger d - m^2) \star H &=& 2m (d^\dagger d - m^2) \star d B = 2m \star d (d^\dagger d - m^2) B = 2m \star d (-2m \star dA)
\nonumber \\
&=& - 4 m^2 d^\dagger d A.
\end{eqnarray}
Thus, it is easy to see that the eom for the $A$ field can be written as
\begin{equation}
( d^\dagger d + m^2)^2 A = 0.
\end{equation}
At the same time, we can perform similar manipulations to eq. \eqref{eom_A2} to arrive at the equation
\begin{equation}\label{EOM_2'}
( d^\dagger d + m^2)^2 B = 0.
\end{equation}

In the next section we analyze even more these resulting equations and their classes of solutions. We finish this section
observing that if we apply again the exterior derivative we arrive at equations over the strength fields
\begin{align}
\label{eom_F} (\square + m^2)^2 F &= 0, \\
\label{eom_H} (\square + m^2)^2 H &= 0,
\end{align}
we have used that $F$ and $H$ are closed forms.

%%%%%%%%%%%%%%%%%%%%%%%%%%%%%%%%%%%%%%%%%%%%%%%%%%%%%%%%%%%%%%%%%%%%%%%%%%%%%%%%%%%%%%%%%%%%%%%%%%%%%%%%%%%%%%%%%%%%%%%%%%%%%%%%%%%%%%%%%%%
\section{Homological Analysis of the solutions to the equations of motion}
%%%%%%%%%%%%%%%%%%%%%%%%%%%%%%%%%%%%%%%%%%%%%%%%%%%%%%%%%%%%%%%%%%%%%%%%%%%%%%%%%%%%%%%%%%%%%%%%%%%%%%%%%%%%%%%%%%%%%%%%%%%%%%%%%%%%%%%%%%%

Having the equations of motion for the potentials, we analyze the characteristics of the 
solutions in order to having a way to describe the manifold under the fields evolve.

As we observed in the previous section, the solutions for $F$ and $H$ satisfy the square of 
the Klein-Gordon like (Proca) equation,
\begin{align} \label{kgA2}
    \left( \square + m^2 \right) F &= 0,\\ \label{kgB2}
    \left( \square + m^2 \right) H &= 0,
\end{align}
note that these automatically satisfy the eom as well, i.e., equations (\ref{eom_F}) and (\ref{eom_H}), respectively.

These equations after we factoring a `$d$' can be written in terms of the potentials, be equal to
\begin{align} \label{kgA3}
    \left( d^{\dagger} d + m^2 \right) A &= \alpha',\\ \label{kgB3}
    \left( d^{\dagger} d + m^2 \right) B &= \beta',
\end{align}
where $\alpha'$ and $\beta'$ are elements in the first and second cohomology classes of $M_4$, respectively. 
Considering equation (\ref{kgB3}) in the next form $d^{\dagger} d B = \beta' - m^2 B$, substituting into \eqref{eom_A2}, for
a closed 2-form $\alpha$, we get
\begin{equation}\label{alpha}
   \star dA = m B + \frac{1}{2m} (\alpha - \beta').
\end{equation}
In a procedure completely analogous, using now \eqref{kgA3} and substituting into \eqref{eom_B2}, for a closed 1-form $\beta$,
we have
\begin{equation} \label{beta}
   \star dB = - m A + \frac{1}{2m} (\alpha' - \beta).
\end{equation}
Equations \eqref{alpha} and \eqref{beta} are a set of relations of linear equations between the fields an their field strength, 
it is worth to mention that these relations are cohomology classes since they are tied to class. In terms of their corresponding
field strength,
\begin{align}\label{eq_FA}
    \star F &= m B + \frac{1}{2m} (\alpha - \beta'),\\ \label{eq_FB}
    \star H &= -m A + \frac{1}{2m} (\alpha' - \beta).
\end{align}
These later expressions (\ref{eq_FA}) and (\ref{eq_FB}), establish a linear relation to compute $A$ in terms of $B$, or conversely 
$B$ in terms $A$. Moreover, they are source of each other \footnote{Thus the field $B$ is source of the field $A$, and
viceversa, with a of topological origin.}. These equations can be regarded ``analogously" to Maxwell equations, by applying 
$\star d$ on mentioned equations, obtaining
\begin{align}\label{eq_FA_J}
     d^\dagger F &= m J_F,\\ \label{eq_FB_J}
     d^\dagger H &= -m J_H,
\end{align}
where $J_F = \star H$ and $J_H = \star F$ can be regarded as currents density, thus the fields strength $F$ and $H$ are 
mutually sources of each other.

Within the present class of solutions, we can perform a gauge transformation \eqref{gaugeA} and \eqref{gaugeB}, and applying
$d^\dagger$ into eqs. \eqref{eq_FA} and \eqref{eq_FB}, we arrive to the conditions
\begin{align}
    d^\dagger A & = 0, \label{coho_A} \\
    d^\dagger B & = 0.  \label{coho_B}
\end{align}
As we can see, they're are co-closed forms, can be regarded as continuity equations. Hence, $\star B$ belongs to the second cohomology group, while $\star A$ to the 
third cohomology group. In both cases the solutions they might be either homologically trivial or not.  

We conclude that the equations for the fields $F$ and $H$, are equivalent to solve the set of equations $(\ref{eq_FA})$ and $(\ref{eq_FB})$, subject to the restriction $(\ref{coho_A})$ and $(\ref{coho_B})$.

%%%%%%%%%%%%%%%%%%%%%%%%%%%%%%%%%%%%%%%%%%%%%%%%%%%%%%%%%%%%%%%%%%%%%%%%%%%%%%%%%%%%%%%%%%%%%%%%%%%%%%%%%%%%%%%%%%%%%%%%%%%%%%%%%%%%%%%%%%%
\subsection{Solution for $(dd^{\dagger}+4m_0^2k^2)^2 F, H = 0$}
%%%%%%%%%%%%%%%%%%%%%%%%%%%%%%%%%%%%%%%%%%%%%%%%%%%%%%%%%%%%%%%%%%%%%%%%%%%%%%%%%%%%%%%%%%%%%%%%%%%%%%%%%%%%%%%%%%%%%%%%%%%%%%%%%%%%%%%%%%%

We finalize this section be studying what topological information the solutions to eqs \eqref{eom_F} and \eqref{eom_H} have. Let us concentrate in
eq. \eqref{eom_F} for now. We can have in general that
\begin{eqnarray}\label{eq_AG}
 (dd^{\dagger} + m^2 ) F = M, 
\end{eqnarray}
for some 2-form $M$. Also we can check explicitly by applying again $d$ to the previous equation that $M$ belongs to the second cohomology 
group. Thus, we again look for solutions to the equation
\begin{equation}\label{eq_AG_2}
(dd^\dagger + m^2) M = 0,
\end{equation}
for a closed 2-form as we see previously in \eqref{kgA2}. However, looking closely into eq. \eqref{eq_AG} gives us that $M$
is in fact cohomologically trivial, due to
\begin{equation}
d ( d^\dagger d + m^2 ) A = M.
\end{equation}
Hence, from the topological point of view, the non-trivial solutions are related with the solutions for $(d^{\dagger}d+m^2)A$ which is a closed form, mean while the solutions  $(d^{\dagger}d+m^2)^2 A$ are trivial solutions. Then, solutions reduce just to those that we analyze in the previous subsection.

At the same time we have for the $B$ field,
\begin{align}\label{eq_BH}
 (dd^\dagger + m^2 ) H &= N,\\
\label{eq_AG_2}
(dd^\dagger + m^2 ) N &= 0,
\end{align}
for a 3-form $N$, which is also an element of $H^3 (M_4, \mathbb{Z})$, and also cohomologically trivial.

%%%%%%%%%%%%%%%%%%%%%%%%%%%%%%%%%%%%%%%%%%%%%%%%%%%%%%%%%%%%%%%%%%%%%%%%%%%%%%%%%%%%%%%%%%%%%%%%%%%%%%%%%%%%%%%%%%%%%%%%%%%%%%%%%%%%%%%%%%%
\section{The case for Minkowski spacetime}
%%%%%%%%%%%%%%%%%%%%%%%%%%%%%%%%%%%%%%%%%%%%%%%%%%%%%%%%%%%%%%%%%%%%%%%%%%%%%%%%%%%%%%%%%%%%%%%%%%%%%%%%%%%%%%%%%%%%%%%%%%%%%%%%%%%%%%%%%%%

As an example we solve the equations of motion for the Minkowski vacuum, assuming that all the fields only depend on 
time $t$, assuming that the metric have the following form
\begin{equation}
    ds^2 = dt^2 - dx^2 - dy^2 - dz^2.
\end{equation}
After a (large) gauge transformation, the relation \eqref{alpha} between the fields yield
\begin{subequations}\label{MA}
\begin{align}
    A_0 &= 0, \\
    A_1 &= -\frac{1}{m} \partial_0 B_{23},\\
    A_2 &= \frac{1}{m} \partial_0B_{13},\\
    A_3 &= -\frac{1}{m} \partial_0 B_{12},
\end{align}
\end{subequations}
and also from \eqref{beta},
\begin{subequations}\label{MB}
\begin{align}
    B_{01} &= B_{02} = B_{03} = 0,\\
    B_{12} &= -\frac{1}{m}\partial_0 A_3, \\
    B_{13} &= \frac{1}{m}\partial_0 A_2, \\
    B_{23} &= - \frac{1}{m} \partial_0 A_1.
\end{align}
\end{subequations}
We are going to take these equations as consistency conditions. In order to find the solutions, we use the previous decouple equations
\eqref{kgA3},
\begin{align}
    \partial_0^2 A_i + m^2 A_i &= 0, \qquad i = 1, \dots, 3.
%    \partial_0^2 A_2 - m^2 A_2 &= 0,  \\
%    \partial_0^2 A_3 - m^2 A_3 &= 0.
\end{align}
The solutions to theses equations are readily given by
\begin{align}
    A_i (t) &= a_{i+} \exp \left( i m t \right) + a_{i-} \exp \left( - i m t \right),
%    A_2 (t)= a_{2+} \exp \left( 2m_0kt\right)+a_{2-} \exp \left( -2m_0kt\right),\\
%   A_3 (t)= a_{3+} \exp \left( 2m_0kt\right)+a_{3-} \exp \left( -2m_0kt\right).
\end{align}
where $a_{i+}$ and $a_{i-}$ are constants.

Similarly, for the $B$ field we have eqs. \eqref{kgB3}, in this case they are
\begin{align}
    \partial_0^2 B_{12} + m^2 B_{12} &= 0,
%    \partial_0^2 B_{13}-4m_0^2 k^2 B_{13}=0,  \\
%    \partial_0^2 B_{23}-4m_0^2 k^2 B_{23}=0.
\end{align}
and the same expressions for $B_{13}$ and $B_{23}$. Solutions to these equations are readily given by
\begin{align}
    B_{12} (t) &= b_{1+} \exp \left(  i m t \right) + b_{1-} \exp \left( - i m t\right),
%    B_{13} (t)= b_{2+} \exp \left( 2m_0kt\right)+b_{2-} \exp \left( -2m_0kt\right),\\
%    B_{23} (t)= b_{3+} \exp \left( 2m_0kt\right)+b_{3-} \exp \left( -2m_0kt\right).
\end{align}
and similar expressions for $B_{13}$ and $B_{23}$; $b_{i+}$ and $b_{i-}$ are constants. Its worth noting that the time
dependance of these fields look like the non-relativistic solutions to the Klein-Gordon equation.

In order that relations \eqref{MA} and \eqref{MB} are fulfilled the coefficients must satisfy, for instance,
\begin{equation}
a_{3+} = - i b_{1+}, \qquad a_{3-} = i b_{1-}.
\end{equation}

A last observation is that eqs. \eqref{MA} and \eqref{MB} show a kind of duality between the fields, since the `magnetic' 
part of a field strength are given by the `electric' part of the correspondent other field strength, as we can obtain from the
original equations of motion.

%%%%%%%%%%%%%%%%%%%%%%%%%%%%%%%%%%%%%%%%%%%%%%%%%%%%%%%%%%%%%%%%%%%%%%%%%%%%%%%%%%%%%%%%%%%%%%%%%%%%%%%%%%%%%%%%%%%%%%%%%%%%%%%%%%%%%%%%%%%
\section{Solutions for Schwarzschild spacetime}
%%%%%%%%%%%%%%%%%%%%%%%%%%%%%%%%%%%%%%%%%%%%%%%%%%%%%%%%%%%%%%%%%%%%%%%%%%%%%%%%%%%%%%%%%%%%%%%%%%%%%%%%%%%%%%%%%%%%%%%%%%%%%%%%%%%%%%%%%%%

As our second and final example we calculate an equatorial solution ($\theta=x^2=\pi/2$  therefore $dx^2 =  0$), for this model within a Schwarzschild background. First, we write down explicitly the eqs. \eqref{kgA3}
and \eqref{kgB3} for the field $A$ and $B$ assuming that their components are only functions of $t$ and $r$. Explicitly these equations read for $A$
\begin{eqnarray}\label{eq:eomA1}
    -\frac{1}{r^2}\left( 1-\frac{2M}{r} \right) \frac{\partial}{\partial r}\left(r^2 \left (\frac{\partial}{\partial t}A_1-\frac{\partial}{\partial r}A_0 \right) \right) + m^2 A_0 &=& 0, \\
    \label{eq:eomA2}
    -\frac{1}{r^2}\left(1-\frac{2M}{r} \right)^{-1} \frac{\partial}{\partial t}\left(r^2 \left(\frac{\partial}{\partial t}A_1-\frac{\partial}{\partial r}A_0 \right) \right) + m^2 A_1 &=& 0, \\
    %\label{eq:eomA2}
    %-\frac{\partial}{\partial t} \left[ \left(1-\frac{2M}{r}\right)^{-1} \frac{\partial}{\partial t} A_2 \right] - \frac{\partial}{\partial r} \left[\left( 1-\frac{2M}{r} \right) \frac{\partial}{\partial r}A_2 \right] +4 m_0^2 k^2=0\\
    \label{eq:eomA3}
    -\frac{\partial}{\partial t}\left[ \left( 1-\frac{2M}{r}\right)^{-1} \frac{\partial}{\partial t}A_3\right] +\frac{\partial}{\partial r}\left[  \left( 1-\frac{2M}{r}\right)  \frac{\partial}{\partial r}A_3 \right] + m^2 A_3 &=& 0,
\end{eqnarray}
and for $B$
\begin{eqnarray}\label{eq:eomB12}
    -\left(1-\frac{2M}{r} \right)^{-1} \frac{\partial}{\partial t}\left(\frac{\partial}{\partial t}B_{12}-\frac{\partial}{\partial r}B_{02} \right) + m^2 B_{12} &=& 0,\\
    \label{eq:eomB13}
    \left(1-\frac{2M}{r} \right)^{-1} \frac{\partial}{\partial t}\left(\frac{\partial}{\partial r}B_{03}-\frac{\partial}{\partial t}B_{13} \right) + m^2 B_{13} &=& 0,\\
    \label{eq:eomB23}
    -\frac{\partial}{\partial t} \left[\left( \frac{1}{r^2} \right)\left(1-\frac{2M}{r} \right)^{-1} \frac{\partial}{\partial t}B_{23} \right] +\frac{\partial}{\partial r} \left[\left( \frac{1}{r^2} \right)\left(1-\frac{2M}{r} \right) \frac{\partial}{\partial r}B_{23} \right] + m^2 B_{23}
    &=& 0,\\
    \label{eq:eomB02}
     -\left(1-\frac{2M}{r} \right) \frac{\partial}{\partial r}\left(\frac{\partial}{\partial t}B_{12}-\frac{\partial}{\partial r}B_{02} \right) + m^2 B_{02} &=& 0,\\
     \label{eq:eomB03}
      -\left(1-\frac{2M}{r} \right)^{-1} \frac{\partial}{\partial r}\left(\frac{\partial}{\partial r}B_{03}-\frac{\partial}{\partial t}B_{13} \right) + m^2 B_{03} &=& 0,
\end{eqnarray}
where we consider the metric $g_{\mu \nu}=diag((1-\frac{2M}{r}),-(1-\frac{2M}{r})^{-1},-r^2,-r^2)$ (we use $\sin(\pi/2)=1$). Considering solutions $A_0$ and $A_1$, can be solved from the first two equation ($\ref{eq:eomA1}$) and ($\ref{eq:eomA2}$), but in order to decouple the equations we must increase the derivatives order and the number of terms with derivatives and similarly for the components $B_{02}$, $B_{03}$, $B_{12}$ and $B_{13}$; rather we use the equations ($\ref{beta}$) and ($\ref{alpha}$) that in components give the following equations for($\ref{beta}$)
\begin{eqnarray}
    \label{eq:beta1}
    A_3 &=& \frac{1}{m}\left[\frac{\partial}{\partial r}B_{02}-\frac{\partial}{\partial t}B_{12}\right],\\
    \label{eq:beta2}
    A_1 &= &-\frac{1}{m r^2} \left(1 -\frac{2M}{r}\right)^{-1}\frac{\partial}{\partial t}B_{23},\\
    \label{eq:beta3}
A_0 &=& \frac{1}{m r^2} \left(1-\frac{2M}{r}\right)\frac{\partial}{\partial r}B_{23},
\end{eqnarray}
and for ($\ref{alpha}$) we find
\begin{eqnarray}\label{eq:alfa01}
    B_{01} &=& 0,\\
    \label{eq:alfa03}
    B_{03} &=& \frac{1}{m}\left(1- \frac{2M}{r} \right) \frac{\partial}{\partial r} A_2 ,\\
    \label{eq:alfa13}
    B_{13} &=& \frac{1}{m}\left(1- \frac{2M}{r} \right)^{-1} \frac{\partial}{\partial t} A_2,
\end{eqnarray}
These equations will be useful to find relations in order to simplify the computations. Considering equation ($\ref{eq:beta1}$) and substituting into (\ref{eq:eomB12}), (\ref{eq:eomB02}), we get the relations
\begin{eqnarray}\label{exp_B12}
    B_{12} &=& -\frac{1}{m}\left(1 -\frac{2M}{r} \right)^{-1} \frac{\partial}{\partial t} A_3,\\\label{exp_B02}
    B_{02} &=& -\frac{1}{m}\left(1 -\frac{2M}{r} \right) \frac{\partial}{\partial r} A_3.
\end{eqnarray}
Taking equation $(\ref{eq:eomB23})$, and substituing $(\ref{eq:beta2})$ and $(\ref{eq:beta3})$ we find
\begin{equation}\label{exp_B23}
    B_{23}=-\frac{1}{m}\left( \frac{\partial}{\partial t}A_1 + \frac{\partial}{\partial r}A_0 \right).
\end{equation}
In order to be consistent and get the same number of degrees of freedom for both fields; because $A_2$ does not appear since we assume $x_2=constant$ then the components $B_{03}=B_{13}=0$. Therefore we can see in our relations that the equation $A_0$ and $A_1$ are related with $B_{23}$, similarly $B_{02}$ and $B_{12}$ a related to $A_3$.
%Now consider ($\ref{eq:alfa03}$) y ($\ref{eq:alfa13}$) and substituting into ($\ref{eq:eomB13}$) we obtain the following equation for the component field $A_2$

%%%%%%%%%%%%%%%%%%%%%%%%%%%%%%%%%%%%%%%%%%%%%%%%%%%%%%%%%%%%%%%%%%%%%%%%%%%%%%%%%%%%%%%%%%%%%%%%%%%%%%%%%%%%%%%%%%%%%%%%%%%%%%%%%%%%%%%%%%%
\subsection{Solving the equations}
%%%%%%%%%%%%%%%%%%%%%%%%%%%%%%%%%%%%%%%%%%%%%%%%%%%%%%%%%%%%%%%%%%%%%%%%%%%%%%%%%%%%%%%%%%%%%%%%%%%%%%%%%%%%%%%%%%%%%%%%%%%%%%%%%%%%%%%%%%%

We now proceed to solve the equations using the next relations from the results (\ref{eq:beta2}) and (\ref{eq:beta3}). Taking the derivatives wrt $r$ and $t$ in order to eliminate $B_{23}$ we obtain the next relation of first derivatives of $A_0$ and $A_1$,
\begin{equation}
    \frac{\partial}{\partial r}\left( r^2 \left( 1- \frac{2M}{r}\right)A_1\right)=r^2 \left( 1- \frac{2M}{r}\right)^{-1} \frac{\partial}{\partial t}A_0.
\end{equation}
Using the method of separation of variables, consider $A_0(t,r)=T_0(t)R_0(r)$ and $A_1(t,r)=T_1(t)R_1(r)$ give rise to the next two useful equations:
\begin{eqnarray}\label{eq:T10}
\frac{1}{T_1}\frac{d}{d t}T_0 &=& C, \\ \label{eq:R10}
R_0&=&\frac{1}{C}\frac{1}{r^2}\left(1-\frac{2M}{r} \right)\frac{d}{dr}\left( r^2 \left(1-\frac{2M}{r} \right)R_1\right),
\end{eqnarray}
where $C$ is a separation constant. Considering the separation of variables ansatz for equation (\ref{eq:eomA2}) (we consider this equation instead of (\ref{eq:eomA1}), since it is slightly easier to manipulate), we obtain the equation
\begin{eqnarray}
    -R_1\frac{d^2}{dt^2}T_1 + \left(\frac{d}{d t} T_0 \right)\left(\frac{d}{d r} R_0\right) + m^2 \left( 1-\frac{2M}{r} \right)T_1 R_1=0,
\end{eqnarray}
where, with the aid of (\ref{eq:T10}) and (\ref{eq:R10}) we replace $\frac{d}{dt}T_0$ and $R_0$ into the second term, we obtain an equation that depends only on $R_1$ and $T_1$,
\begin{equation}\label{eq:eomR1T1}
    -\frac{1}{T_1}\frac{d^2}{dt^2}T_1+\frac{1}{R_1}\frac{d}{d r}\left[ \frac{1}{r^2}\left(1-\frac{2M}{r}\right) \frac{d}{d r}\left( r^2\left( 1-\frac{2M}{r} \right) R_1\right) \right] + m^2 \left( 1-\frac{2M}{r} \right)=0.
\end{equation}
As we can check explicitly, it can be solved by separation of variables, 
\begin{equation}
    \frac{1}{T_1} \frac{d^2}{d t^2}T_1= \ell^2,
\end{equation}
being $\ell^2$ a separation constant, whose solution is 
\begin{equation}\label{solT1}
    T_1(t)= a_1^+e^{\ell t}+a_1^-e^{-\ell t}.
\end{equation}
Thus the equation (\ref{eq:eomR1T1}) reads
\begin{equation}\label{eq:eomR1}
   \frac{d}{d r}\left[ \frac{1}{r^2}\left(1-\frac{2M}{r}\right) \frac{d}{d r}\left( r^2\left( 1-\frac{2M}{r} \right) R_1\right) \right] + \left[ m^2 \left( 1-\frac{2M}{r} \right) - \ell^2 \right]R_1=0.
\end{equation}
Rewritting
\begin{eqnarray}
    (r^4-4Mr^3+4M^2r^2)\,R_1''+(2r^3-2Mr^2-4M^2r)\,R_1' +\nonumber \\+\Big[( m^2 -\ell^2) r^4 - 2 M m^2 r^3 - 2 r^2 + 12M r - 12 M^2 \Big] R_1 = 0,  
\end{eqnarray}
whose solution in series around $r=2M$ is the following:
\begin{equation}
R_1(r)=  \sum_{j=0}^3 \alpha^+_j (r-2M)^{j+2M\ell-1}  + \sum_{j=0}^3
\alpha^-_j (r-2M)^{j-2M\ell-1},
\end{equation}
where the coefficients $\alpha_j^+$ and $\alpha_j^-$ are explicitly calculated up to third order into Appendix  \ref{Appendix_B}.

Next, to solve $A_0$ we use the relations  (\ref{eq:T10}) and (\ref{eq:R10}), between the fields components $A_1$ and $A_0$. The solution for the temporal component $T_0(t)$ is easily computed integrating respect $t$ in virtue of (\ref{eq:T10}) and the solution (\ref{solT1})
\begin{equation}
    T_0 (t) = \frac{C}{\ell} \big( a_1^+e^{\ell t}-a_1^-e^{-\ell t}\big),
\end{equation}
and for the radial component up to third order, using (\ref{eq:R10}), we get
\begin{eqnarray}  
R_0(r) = \frac{(r-2M)^{2M\ell}}{Cr^2} \sum_{j=0}^{3} \alpha_j^+ (r-2M)^j \left[ 2M\ell+j-1 + \frac{2(r-M)}{r} \right]  \nonumber\\  
+ \frac{(r-2M)^{-2M\ell}}{Cr^2} \sum_{j=0}^{3} \alpha_j^- (r-2M)^j \left[ j-2M\ell-1 + \frac{2(r-M)}{r} \right]. 
\end{eqnarray}
It is worth to point out some considerations: the solutions in the appendixes we use the change of variable $r=x+2M$ in order to simplify the calculations. In addition, the solution $A_0$ is independent of the separation constant $C$, because in the general solution are multiplied, therefore will be canceled out. And finally, the explicit computation can be checked in Appendix \ref{Appendix_C}, bearing in mind that the $\alpha_i^+$ and $\alpha_i^+$, are the same of the solution for $A_1$.

Seeking for solutions of the component $A_3$ use separation variables, hence we assume that $A_3(t,r)=T_3(t)R_3(r)$, then we obtaining the following to equations, for the temporal component
\begin{equation}
    \frac{1}{T_3} \frac{d^2}{dt^2}T_3=n^2,
\end{equation}
where $n$ is the separation constant and the solution is
\begin{equation}
    T_3(t)=D_+ e^{nt}+D_-e^{-nt}.
\end{equation}
For the radial part we have the following equation,
\begin{equation}
    \left(1-\frac{2M}{r}\right)\frac{d}{dr}\left[\left(1-\frac{2M}{r}\right)\frac{d}{dr}R_3\right] +\left[ m^2 \left(1-\frac{2M}{r} \right) -n^2\right]R_3=0. \label{eq:original}
\end{equation}
Next, we seek for solutions to this equation. In order to find it we rewrite the equation as follows.
\begin{equation} 
r(r-2M)^2\frac{d^2R_3}{dr^2} + 2M (r-2M)\frac{dR_3}{dr} + \left[ m^2 (r-2M)r^2 - n^2r^3 \right] R_3 =0.
\label{eq:rewritten-r}
\end{equation}
Solving this expression through Frobenius method (check Appendix \ref{Appendix_D}), we find
\begin{eqnarray} 
    R_3(r) =  \sum_{j=0}^3 \beta_j^+(r-2M)^{j+2Mn}  + \sum_{j=0}^3 \beta_j^-(r-2M)^{j-2Mn}.    
\end{eqnarray}

\subsection{Solutions for the field $B$}

Using the relations $(\ref{exp_B12})-(\ref{exp_B23})$ among the components of $A$ and $B$ we easily find the solutions differentiating 
\begin{eqnarray}
    B_{02} &=& - \frac{1}{m} \left( 1-\frac{2M}{r}\right)T_3(t)\frac{\partial}{\partial r}R_3(r) \nonumber \\ 
   &=&-\frac{1}{m r} \left[D_+ e^{nt}+D_-e^{-nt}\right] \left[ \sum_{j=0}^{3} (j+2Mn)\,\beta_j^+\,(r-2M)^{j+2Mn}+\sum_{j=0}^{3} (j-2Mn)\,\beta_j^-\,(r-2M)^{j-2Mn} \right], \nonumber \\ \\
    B_{12} &=&  -\frac{1}{m}\left(1 -\frac{2M}{r} \right)^{-1} R_3(r) \left(  \frac{\partial}{\partial t}T_3(t)\right)\nonumber \\
    &=& -\frac{nr}{m}\left[D_+ e^{nt} -D_- e^{-nt} \right] \left[ \sum_{j=0}^3 \beta_j^+(r-2M)^{j+2Mn-1}  + \sum_{j=0}^3 \beta_j^-(r-2M)^{j-2Mn-1}\right],\\
    B_{23} &=& -\frac{1}{m}\left( R_1(r)\frac{\partial}{\partial t}T_1(t) + T_0(t)\frac{\partial}{\partial r}R_0(r) \right) \nonumber \\
    &=&-\frac{1}{m} \left[a_1^+e^{\ell t}-a_1^-e^{-\ell t} \right] \cdot \nonumber \\ 
    &\cdot&  \left[\sum_{j=0}^3 \alpha^+_j \left( \ell -\frac{1}{\ell}F_n^{+}(r) \right) (r-2M)^{j+2M\ell-1} + \sum_{j=0}^3
    \alpha^-_j \left( \ell -\frac{1}{\ell}F_n^{-}(r) \right) (r-2M)^{j-2M\ell-1} \right], 
\end{eqnarray}
where
\begin{eqnarray} 
F_n^{\pm}(r) &=& \frac{(n\pm 2M\ell)}{r^3} \Big[r(n+1\pm 2M\ell)-2M\Big] \nonumber\\ &&\quad -\frac{3(r-2M)}{r^4} \Big[r(n+1\pm 2M\ell)-2M\Big] +\frac{(n+1\pm 2M\ell)(r-2M)}{r^3}. 
\end{eqnarray}
As we can see the components of $B_{02}$ and $B_{12}$ depends on only of the mode $n$, while $B_{23}$ depends of both modes $n$ and $\ell$.

%%%%%%%%%%%%%%%%%%%%%%%%%%%%%%%%%%%%%%%%%%%%%%%%%%%%%%%%%%%%%%%%%%%%%%%%%%%%%%%%%%%%%%%%%%%%%%%%%%%%%%%%%%%%%%%%%%%%%%%%%%%%%%%%%%%%%%%%%%%
\section{Conclusions}
\label{Conclusions}
%%%%%%%%%%%%%%%%%%%%%%%%%%%%%%%%%%%%%%%%%%%%%%%%%%%%%%%%%%%%%%%%%%%%%%%%%%%%%%%%%%%%%%%%%%%%%%%%%%%%%%%%%%%%%%%%%%%%%%%%%%%%%%%%%%%%%%%%%%%

Topological methods in physics have been a powerful tool for analyzing and better understanding certain systems. 
In the absence of dynamical degrees of freedom, mass can be topologically generated for fields on a given manifold—as in topologically massive 
gravity. In the case studied in this paper, we have a similar scenario for gauge fields, which are known to lack a mass term due to gauge invariance. 
Using a 
topological $BF$ theory defined on a 7-dimensional manifold, we generate a mass for the two gauge fields $A$ and $B$
by dimensional reduction. One difference we observe compared to the usual (3$d$) $BF$ theory is that the (reduced) $B$ field is no longer
auxiliary but a true dynamical degree of freedom.

Furthermore, we show that analyzing the topological information encoded in the equations of motion allows us to reconstruct characteristics of the 
underlying manifold, provided its cohomology groups are known. Finally, we derive explicit solutions in both Minkowski and Schwarzschild spacetimes. 
Ongoing work focuses on comparing these results with standard classical gauge field solutions in the same backgrounds. Notably, the mutual sourcing 
between the gauge fields hints at an intriguing duality worth pursuing; results along these lines will be reported in a future publication.

\section*{Acknowledgments}
RSS and AMM would like to thank the Universidad de Guadalajara for its support through the PROSNII and PFP programs, as well as SECIHTI for
its support during the preparation of this paper. EAR would like to thank SECIHTI (fomerly CONAHCyT) for a master's
degree scholarship.

%\bibliographystyle{Frontiers-Vancouver}
%\bibliographystyle{Frontiers-Vancuver} %  Many Frontiers journals use the Harvard referencing system (Author-date), to find the style and resources for the journal you are submitting to: https://zendesk.frontiersin.org/hc/en-us/articles/360017860337-Frontiers-Reference-Styles-by-Journal. For Humanities and Social Sciences articles please include page numbers in the in-text citations 
%\bibliographystyle{Frontiers-Vancouver} % Many Frontiers journals use the numbered referencing system, to find the style and resources for the journal you are submitting to: https://zendesk.frontiersin.org/hc/en-us/articles/360017860337-Frontiers-Reference-Styles-by-Journal

%\bibliography{Draft_V2/bib_vortex}

%%% Make sure to upload the bib file along with the tex file and PDF
%%% Please see the test.bib file for some examples of references

%%% If you don't add the figures in the LaTeX files, please upload them when submitting the article.
%%% Frontiers will add the figures at the end of the provisional pdf automatically
%%% The use of LaTeX coding to draw Diagrams/Figures/Structures should be avoided. They should be external callouts including graphics.

..........
\appendix

%%%%%%%%%%%%%%%%%%%%%%%%%%%%%%%%%%%%%%%%%%%%%%%%%%
\section{Differential forms}\label{Appendix_A}
%%%%%%%%%%%%%%%%%%%%%%%%%%%%%%%%%%%%%%%%%%%%%%%%%

In this appendix we discuss some conventions and useful definitions of differential forms that we use throughout the article, in order to write down in a coordinate-free fashion the action 
and the EOM in their most general form. We follow the references \cite{Nakahara:2003nw, Frankel:2011}.

A differential $p$-form $\alpha$ over a $n$-dimensional manifold $M$ is a totally skew-symmetric tensor that is written locally as
\begin{equation}\label{p_form}
    \alpha= \frac{1}{p!}\alpha(x)_{\mu_1 \ldots \mu_p} dx^{\mu_1} \wedge \cdots \wedge dx^{\mu_p} \,,
\end{equation}
where the coefficients $\alpha(x)_{\mu_1 \ldots \mu_p}$ are functions of the manifold $M$; the indices $\mu_i$ runs over $1, \ldots, n$. The factors $dx^{\mu_1} \wedge \cdots \wedge dx^{\mu_p}$ 
are the basis of the vector space of $p$-forms in terms of the coordinates $x^\mu$'s; also $\wedge$ is the wedge product, and it represents the operation among the $1$-forms in order to form an 
skew-symmetric object. Finally, we denote this vector space of all $p$ forms by $\Omega^p (M)$.

We introduce the exterior derivative as a map $d: \Omega^p(M) \to \Omega^{p+1}(M)$ by the expression
\begin{equation}
d \alpha=\frac{1}{p!} (\partial_\nu \, \alpha_{\mu_1 \ldots \mu_p}) dx^\nu \wedge dx^{\mu_1} \wedge \cdots \wedge dx^{\mu_p} \,, 
\end{equation}
where $\alpha$ is the $p$-form in \eqref{p_form}.

On the other hand, the `star' Hodge dual map, $\star: \Omega^p \rightarrow \Omega^{n-p}$, is defined by assigning the $(n-p)$-form dual to the $p$-form $\alpha$, by the relation
\begin{equation}
    \beta \wedge \star \alpha = \langle \beta, \alpha \rangle d \mathrm{vol},
\end{equation}
where $\langle \cdot, \cdot \rangle$ is the pointwise inner product in $\Omega^p (M)$, and $d \mathrm{vol}$ is the invariante riemannian measure. In coordinates, it has the expression
\begin{eqnarray}\label{hodge}
    \star \alpha &:=& \star\left(\frac{1}{p!}\alpha_{\mu_{1}\cdots \mu_{p}}dx^{\mu_1}\wedge \cdots \wedge dx^{\mu_p} \right) \nonumber \\
    &=& \frac{\sqrt{-g}}{p!(n-p)!}\alpha_{\mu_{1}\cdots \mu_p}\epsilon^{\mu_1 \cdots \mu_p}{}_{\nu_1\cdots \nu_{n-p}}dx^{\nu_1}\wedge \cdots \wedge dx^{\nu_{n-p}} \,,
\end{eqnarray}
where $\epsilon^{\mu_1\cdots \mu_p}{}_{\nu_1 \cdots \nu_{n-p}}=g^{\mu_1\rho_1}\cdots g^{\mu_p\rho_p}\epsilon_{\rho_1 \cdots \rho_p \nu_1 \cdots \nu_{n-p}}$, here $n$ is the dimension. For the Lorentzian case, the following property holds for the Hodge star operator
\begin{eqnarray}
    \star^2 \alpha &=& (-1)^{1+p(n-p)}\alpha \,.
\end{eqnarray}

Since, by definition of the Hodge dual, $\beta \wedge \star \alpha$ is proportional to the volume form, hence, we can define a global inner product by the relation
\begin{equation}
(\alpha,\beta):=\int_{M_4}\alpha \wedge \star \beta \,,
\end{equation}\label{effective_action}
with $\alpha, \beta \in\Omega^{p}(M)$. Due to the linearity of $d$ and the inner product, we define its adjoint, $d^\dagger$, by the expression
\begin{equation}
      (d\alpha,\beta)=(\alpha,d^{\dagger}\beta) \,,
\end{equation}
up to a surface term, where $d^\dagger$ has the expression
\begin{equation}\label{dag}
    d^{\dagger} = (-1)^{n(p+1)}\star d\star \,.
\end{equation}

\section{Solution $R_1$}\label{Appendix_B}
%\section*{Frobenius Solution Around $x=0$}

Consider the differential equation:
\begin{eqnarray}
    (r^4-4Mr^3+4M^2r^2)\,R_1''+(2r^3-2Mr^2-4M^2r)\,R_1' +\nonumber \\+\Big[(4m_0^2k^2-\ell^2)r^4-8Mm_0^2k^2r^3-2r^2+12Mr-12M^2\Big]R_1=0.  
\end{eqnarray}
introducing the change of variable

\begin{equation}
  r=x+2M.  
\end{equation}
The point $x=0$ corresponds to the horizon $r=2M$.

\begin{equation}
A(x)R_1''(x) + B(x)R_1'(x)+ C(x)R_1(x) = 0,
\end{equation}

where
\begin{eqnarray}
    A(x)&=&4M^2x^2+4Mx^3+x^4= a_2 x^2+ a_3 x^3+a_4 x^4 \\
    B(x)&=& 12M^2x+10Mx^2+2x^3= b_1 x+ b_2 x^2+b_3 x^3\\
    C(x) &=& (4m_0^2k^2-\ell^2)x^4
+(24Mm_0^2k^2-8M\ell^2)x^3
+(48M^2m_0^2k^2-24M^2\ell^2-2)x^2 \\
&+&(32M^3m_0^2k^2-32M^3\ell^2+4M)x
+\left(4M^2-16M^4\ell^2\right)\\
&=& c_0+ c_1 x + c_2 x^2 + c_3 x^3 + c_4 x^4.
\end{eqnarray}
Here we define the $a_i$, $b_i$ and $c_i$ their respective coefficient $i$ corresponding to the power $x^i$, therefore

\begin{eqnarray}
    (a_2 x^2+ a_3 x^3+ a_4 x^4)R_1'' +
(b_1x+b_2x^2+b_3x^3)R_1' + (c_0+c_1x+c_2x^2+c_3x^3+c_4x^4)R_1 =0.
\end{eqnarray}
Now consider the Frobenius Ansatz

\begin{equation}
R_1(x)=\sum_{j=0}^{\infty} \alpha_j x^{j+s},
\qquad \alpha_0\neq 0.
\end{equation}
The derivatives are

\begin{align}
R_1'(x)
&=
\sum_{n=0}^{\infty}
(j+s) \alpha_j x^{j+s-1},
\\
R_1''(x)
&=
\sum_{j=0}^{\infty}
(j+s)(j+s-1) \alpha_j x^{j+s-2}.
\end{align}
The indicial equation close to $x=0$, after some simple algebra

\begin{equation}
s(s-1)+3s+1-4M^2\ell^2=0.
\end{equation}
whose roots are

\begin{equation}
s_+=-1+2M\ell, \qquad s_-=-1-2M\ell.
\end{equation}

The general recurrence relation for $n \geq 4$ after collecting equal powers of $x$, one obtains

\begin{equation}
A_j \alpha_j + B_j \alpha_{j-1} + D_j \alpha_{j-2} + c_3 \alpha_{j-3} + c_4 \alpha_{j-4}=0,
\end{equation}
where

\begin{eqnarray}
A_j &=& 4M^2\Big[(j+s+1)^2-(2M\ell)^2\Big], \\
B_j &=&  4M(j+s-1)^2 + 6M(j+s-1) + c_1
\end{eqnarray}

and

\begin{equation}
D_j = (j+s-2)^2+(j+s-2)+c_3.
\end{equation}
The recurrence formula is the next expression (for $j \geq 4$)

\begin{equation}
\alpha_j= -\frac{ B_j \alpha_{j-1} +
D_j \alpha_{j-2} + c_3 \alpha_{j-3} + c_4 \alpha_{j-4}}{A_j},
\end{equation}
Now we write down the expressions for the first, second and third order:

\begin{eqnarray}
 \alpha_1&=&-\frac{B_1}{A_1}\alpha_0,\\
 \alpha_2&=&-\frac{B_2 \alpha_1+ D_2 \alpha_0}{A_2}=-\frac{-B_2 B_1+D_2A_1}{A_2 A_1} \alpha_0,\\
 \alpha_3&=&-\frac{B_3\alpha_2+ D_3 \alpha_1 + c_3 \alpha_0}{A_3}= -\frac{B_3 B_2 B_1+ B_3 D_2 A_1-D_3 B_1 A_2 +c_3}{A_3 A_2 A_1}\alpha_0.
\end{eqnarray}
We will compute the solutions up to third order for the positive sign of root $s_+=-1+2M\ell$, then substituting into the previous expressions of $\alpha$'s we get

the solution is

\begin{equation}
R_1^+(x)
=
x^{-1+2M\ell}
\sum_{j=0}^{\infty}\alpha^+_j x^j.
\end{equation}

let us compute the solution up to third order. At first order
The recurrence coefficients are 

\begin{eqnarray} 
A_j^{(+)} &=& 4M^2\,j(j+4M\ell),\\
B_j^{(+)} &=& 4M(j+2M\ell-2)^2 + 6M(j+2M\ell-2) + c_1,\\ 
D_j^{(+)} &=& (j+2M\ell-3)^2 + (j+2M\ell-3) + c_2. 
\end{eqnarray} 
The first three coefficients are 

\begin{eqnarray} 
\alpha^+_1 &=& -\frac{ 4M(2M\ell-1)^2 + 6M(2M\ell-1) + c_1} { 4M^2(1+4M\ell) } \,\alpha^+_0.\\ 
 \alpha^+_2 &=& -\frac{ (16M^3\ell^2+12M^2\ell+c_1)\alpha^+_1 + \left[(2M\ell-1)^2+(2M\ell-1)+c_2\right] \alpha^+_0 } { 8M^2(2+4M\ell) }. \\
\alpha^+_3 &=& -\frac{ \left[4M(2M\ell+1)^2+6M(2M\ell+1)+c_1\right]\alpha^+_2 + \left[4M^2\ell^2+2M\ell+c_2\right] \alpha^+_1 + c_3 \alpha^+_0 } { 12M^2(3+4M\ell) }.  
\end{eqnarray}
Now we consider the solution associated to the negative sign of the root $s_-=-1-2M\ell$, the solution is given by

\begin{equation}
R_1^-(x)
=
x^{-1-2M\ell}
\sum_{j=0}^{\infty}\alpha^-_j x^j,
\end{equation}
remember that zeroth order is  $\alpha^-_0$, in order to compute the series after substituting $s_-$ we obtain

\begin{eqnarray} 
A_j^{(-)} &=& 4M^2\,j(j-4M\ell),\\  
B_j^{(-)} &=& 4M(j-2M\ell-2)^2 + 6M(j-2M\ell-2) + c_1,\\
D_j^{(-)} &=& (j-2M\ell-3)^2 + (j-2M\ell-3) + c_2. 
\end{eqnarray}

then the first three coefficients are

\begin{eqnarray} 
\alpha^-_1 &=& -\frac{ 4M(2M\ell+1)^2 + 6M(2M\ell+1) + c_1 } { 4M^2(1-4M\ell) } \,\alpha^-_0.  \\
  \alpha^-_2 &=& -\frac{ (16M^3\ell^2-12M^2\ell+c_1) \alpha^-_1 + \left[(-2M\ell-1)^2+(-2M\ell-1)+c_2\right] \alpha^-_0 }{ 8M^2(2-4M\ell) }.\\  
  \alpha^-_3 &=& - \frac{\left[4M(2M\ell-1)^2 + 6M(2M\ell-1) + c_1\right] \alpha^-_2 + \left[4M^2\ell^2-2M\ell+c_2\right]b\alpha^-_1 + c_3 \alpha^-_0}{ 12M^2(3-4M\ell)}.  
\end{eqnarray}
Finally the general solution up to third order for $R_1(x)$, around $x=0$, i.e., $r=2M$ substituting back in terms of $r$

\begin{equation}
R_1(r)=  \sum_{j=0}^3 \alpha^+_j (r-2M)^{j+2M\ell-1}  + \sum_{j=0}^3
\alpha^-_j (r-2M)^{j-2M\ell-1},
\end{equation}
where the solution is in terms of the constants $\alpha_0^+$ and $\alpha_0^-$.

\section{Solution $R_0$}\label{Appendix_C}
Starting from 
\begin{equation} 
R_0(r) = \frac{1}{C} \frac{1}{r^2} \left( 1-\frac{2M}{r} \right) \frac{d}{dr} \left[ r^2 \left( 1-\frac{2M}{r} \right) R_1(r) \right], 
\end{equation} 
using the change of variable $r=x+2M$, we obtain 

 \begin{equation} 
 R_0(x) = \frac{1}{C} \left[ \frac{x^2}{(x+2M)^2}R_1'(x) + \frac{2x(x+M)}{(x+2M)^3}R_1(x) \right].  
 \end{equation} 
 considering the Frobenius solution for $R_1(x)$, up to third order 
 
\begin{equation}
R_1(x)
= x^{-1+2M\ell}
\left(
\alpha^+_0+\alpha^+_1x+\alpha^+_2 x^2+ \alpha^+_3 x^3 \right) +
x^{-1-2M\ell} \left(
\alpha^-_0 + \alpha^-_1 x + \alpha^-_2 x^2+ \alpha^-_3 x^3 \right),
\end{equation}
differentiating  $R_1^{(+)}$,

\begin{eqnarray} 
R_1^{(+)\prime}(x) = (2M\ell-1)a_0x^{2M\ell-2} +2M\ell a_1x^{2M\ell-1} +(2M\ell+1)a_2x^{2M\ell} +(2M\ell+2)a_3x^{2M\ell+1}. 
\end{eqnarray} 
Substituting into $R_0$ gives 

\begin{eqnarray} 
R_0^{(+)}(x) = \frac{x^{2M\ell}}{C(x+2M)^2} \Bigg[  \alpha_0^+ \left( 2M\ell-1 + \frac{2(x+M)}{(x+2M)} \right) +\alpha_1^+ x \left( 2M\ell + \frac{2(x+M)}{(x+2M)} \right) \nonumber\\  +\alpha_2^+ x^2 \left( 2M\ell+1 + \frac{2(x+M)}{(x+2M)} \right) + \alpha_3^+ x^3 \left( 2M\ell+2 + \frac{2(x+M)}{(x+2M)} \right) \Bigg]. 
\end{eqnarray} 
Similarly for $R_1^{(-)}$

\begin{equation} 
R_1^{(-)\prime}(x) =(-2M\ell-1)b_0x^{-2M\ell-2} -2M\ell b_1x^{-2M\ell-1} +(1-2M\ell)b_2x^{-2M\ell} +(2-2M\ell)b_3x^{1-2M\ell}. 
\end{equation} 
then $R_0$ gives 

\begin{eqnarray} 
R_0^{(-)}(x) = \frac{x^{-2M\ell}}{C(x+2M)^2} \Bigg[  \alpha_0^- \left( -2M\ell-1 + \frac{2(x+M)}{(x+2M)} \right)   + \alpha_1^- x \left( -2M\ell + \frac{2(x+M)}{(x+2M)} \right) \nonumber\\  + \alpha_2^- x^2 \left( 1-2M\ell + \frac{2(x+M)}{(x+2M)} \right)  + \alpha_3^- x^3 \left( 2-2M\ell + \frac{2(x+M)}{(x+2M)} \right) \Bigg]. \end{eqnarray} 
Finally the general solution in terms of $r$ considering both branches is,

\begin{eqnarray}  
R_0(r) = \frac{(r-2M)^{2M\ell}}{Cr^2} \sum_{j=0}^{3} \alpha_j^+ (r-2M)^j \left[ 2M\ell+j-1 + \frac{2(r-M)}{r} \right]  \nonumber\\  
+ \frac{(r-2M)^{-2M\ell}}{Cr^2} \sum_{j=0}^{3} \alpha_j^- (r-2M)^j \left[ j-2M\ell-1 + \frac{2(r-M)}{r} \right]. 
\end{eqnarray}

\section{Solution $R_3$}\label{Appendix_D}

\begin{equation} \left(1-\frac{2M}{r}\right) \frac{d}{dr} \left[ \left(1-\frac{2M}{r}\right) \frac{dR_3}{dr} \right] + \left[ 4m_0^2k^2 \left(1-\frac{2M}{r}\right) -n^2 \right]R_3 =0, \end{equation} 
introducing the change of variable 
 $x=r-2M$, the equation finally takes the following form

\begin{equation} 
x^2(x+2M)\frac{d^2R_3}{dx^2} + 2Mx\frac{dR_3}{dx} + \left[ 4m_0^2k^2\,x(x+2M)^2 - n^2(x+2M)^3 \right] R_3 =0. 
\end{equation} 
The polynomial coefficients are 

\begin{eqnarray} 
A(x)&=&x^3+2Mx^2=a_3 x^3+a_2 x^2,\\ 
B(x)&=&2Mx=b_1 x,\\
C(x)&=&(4m_0^2k^2-n^2)x^3+(16M m_0^2 k^2 - 6Mn^2)x^2 + (16M^2 m_0^2 k^2-12M^2 n^2)x -8M^3n^2 \nonumber \\ &=& c_3x^3+c_2x^2+c_1 x+c_0, 
\end{eqnarray}
then the equation is rewritten as

\begin{equation} 
A(x)\frac{d^2R_3}{dx^2} + B(x)\frac{dR_3}{dx} + C(x)R_3 =0. 
\end{equation} 
We are looking for a series solution around the singular point $x=0$ ($r=2M$). Considering the Frobenius ansatz 

\begin{equation} 
R_3(x) = \sum_{j=0}^{\infty} \beta_j x^{j+s}, \qquad \beta_0\neq0,
\end{equation} 
which derivatives are

\begin{eqnarray} 
R_3'(x) &=& \sum_{j=0}^{\infty} (j+s)\beta_j x^{j+s-1},\\
R_3''(x) &=& \sum_{j=0}^{\infty} (j+s)(j+s-1) \beta_j x^{j+s-2}. 
\end{eqnarray}
Substituting into the differential equation we obtain the expression

\begin{eqnarray} \label{eq:seriesR3}
\sum_{j=0}^{\infty} \left\{  \left[ (j+s)(j+s-1)2M  + (j+s) 2M   -8M^3n^2 \right]\beta_j x^{j+s} \right. \nonumber \\ +\left[(j+s)(j+s-1)+ (16Mm_0^2k^2-6M n^2)  \right] \beta_j x^{j+s+1}+ \nonumber \\
\left.+\left[16Mm_0^2k^2-6Mn^2 \right] \beta_j x^{j+s+2}+ \left[ 4m_0^2k^2-n^2\right] \beta_j x^{j+s+3} \right\}=0.
\end{eqnarray} 
Hence the indicial equation (lowest power) is

\begin{equation} 
\left[ 2Ms(s-1) + 2Ms - 8M^3n^2 \right] a_0. 
\end{equation} 
whose roots are

\begin{eqnarray} 
r_+=2Mn, \\
r_-=-2Mn , 
\end{eqnarray} 
rearranging the indices of equation ($\ref{eq:seriesR3}$) in order to factor out the term $x^{j+s}$, we obtain the following expansion

\begin{eqnarray}
    A_n \beta_n+B_n \beta_{n-1}+C \beta_{n-2}+D \beta_{n-3}= 0,
\end{eqnarray}
for $n\geq3$, where the coefficients are given by the expressions

\begin{eqnarray} \label{coeffA3}
A_j&=& 2Mj(j+2s),  \nonumber \\  \label{coeffB3}
B_j &=& (j+s-1)(j+s-2) + 16M^2m_0^2k^2 - 12M^2n^2,  \nonumber \\  
C &=& 16Mm_0^2k^2-6Mn^2, \nonumber \\  
D &=& 4m_0^2k^2-n^2 . 
\end{eqnarray}
 We have used the indicial equation to reduce the expression $A_j$, then the recurrence formula (for $j \geq 3$)

\begin{equation}
    \beta_j=-\frac{B_j \beta_{j-1}+C \beta_{j-2}+D \beta_{j-3}}{A_j},
\end{equation}
 and the expression for the first and second order are

 \begin{eqnarray}
     \beta_1 &=& -\frac{B_1}{A_1} \beta_0, \\
     \beta_2&=&-\frac{B_2 \beta_1+ C\beta_0}{A_2}=-\frac{-B_2 B_1+C A_1}{A_2 A_1} \beta_0,
 \end{eqnarray}
 assuming that the zero-th order $\alpha_0 \neq 0$.

Up to this point, we will compute the solution until third order for the positive root $s_+=2Mn$, subtituting in to the solution we obtain

\begin{equation}
   R_3^+(x) = x^{2Mn}\sum_{j=0}^{\infty} \beta_j^+ x^{j}, 
\end{equation}
therefore 

\begin{eqnarray}
    A_j^+&=&2Mj(j+4Mn),\\
    B_j^+&=&(j-1+2Mn)(j-2+2Mn)+16M^2m_0^2k^2-12 M^2n^2,
\end{eqnarray}
the expressions of the coefficients up to third order is

\begin{eqnarray}
    \beta_1^+&=&-\frac{(2Mn)(2Mn-1)+16M^2m_0^2k^2-12 M^2n^2}{2M(1+4Mn)}\beta_0^+,\\
    \beta_2^+ &=& -\frac{ \left[(2Mn+1)(2Mn) + 16M^2m_0^2k^2 - 12M^2n^2 \right]\beta_1^+ + (16Mm_0^2k^2-6Mn^2)\beta_0^+ }{ 4M(2Mn+1) } ,\\ 
    \beta_3^+ &=& -\frac{ \left[(2Mn+2)(2Mn+1) + 16M^2m_0^2k^2 - 12M^2n^2 \right]\beta_2^+ + (16Mm_0^2k^2-6Mn^2) \beta_1^+ + (4m_0^2k^2-n^2)\beta_0^+ }{ 6M(4Mn+3) }.  
    \end{eqnarray} 
Now consider the negative root $s_-=-2Mn$, then the solution is

\begin{equation}
   R_3^-(x) = x^{-2Mn}\sum_{j=0}^{\infty} \beta_j^- x^{j}, 
\end{equation}
then the expression (\ref{coeffA3}) and (\ref{coeffB3}) read 

\begin{eqnarray}
    A_j^-&=&2Mj(j-4Mn),\\
    B_j^-&=&(j-1-2Mm)(j-2-2Mn)+16M^2m_0^2k^2-12 M^2n^2,
\end{eqnarray}
in this way we compute up to third order the coefficients

\begin{eqnarray} 
\beta_1^- &=& -\frac{ (-2Mn)(-2Mn-1) + 16M^2m_0^2k^2 - 12M^2n^2 }{ 2M(1-4Mn) } \beta_0^-,  \\ 
\beta_2^- &=& -\frac{ \left[(-2Mn+1)(-2Mn) + 16M^2m_0^2k^2 - 12M^2n^2 \right]\beta_1^- + (16Mm_0^2k^2-6Mn^2)\beta_0^- }{ 4M(1-2Mn) },  \\ 
\beta_3^- &=& -\frac{ \left[(-2Mn+2)(-2Mn+1) + 16M^2m_0^2k^2 - 12M^2n^2 \right] \beta_2^- + (16Mm_0^2k^2-6Mn^2)\beta_1^- + (4m_0^2k^2-n^2) \beta_0^- }{ 6M(3-4Mn) }. 
\end{eqnarray} 
Thus the final solution up to third order in terms of $r$ is 
    
\begin{eqnarray} 
    R_3(r) =  \sum_{j=0}^3 \beta_j^+(r-2M)^{j+2Mn}  + \sum_{j=0}^3 \beta_j^-(r-2M)^{j-2Mn}.    
\end{eqnarray}
%%%%%%%%%%%%%%%%%%%%%%%%%%%%%%%%%%%%%%%%%%%

\bibliography{bib_vortex}
\bibliographystyle{unsrt}
%%%%%%%%%%%%%%%%%%%%%%%%%%%%%%%%%%%%%%%%%%%%%%%%%%%%%%%
%%%%%%%%%%%%%%%%%%%%%%%%%%%%%%%%%%%%%%%%%%%%%%%%%%%%%%%
%%%%%%%%%%%%%%%%%%%%%%%%%%%%%%%%%%%%%%%%%%%%%%%%%%%%%%%
%%%%%%%%%%%%%%%%%%%%%%%%%%%%%%%%%%%%%%%%%%%%%%%%%%%%%%%

\end{document}